# An original way for producing a 2.4 GPa strength ductile steel by rolling of martensite.

J.P. Masse[1], B. Chéhab[2], H. Zurob[3], D. Embury[3], X. Wang[3], O. Bouaziz[1,4]

[1]ArcelorMittal Research and Development Voie Romaine-BP30320, 57283 Maizières-lès-Metz Cédex France

[2]ArcelorMittal Global R&D BP 15, 62330, Isbergues, France

[3]Department of Materials Science and Engineering, McMaster University Hamilton, Ontario, L8S 4L7, Canada

[4]Centre des Matériaux/Mines Paris, Paristech, CNRS-UMR7633, BP 87, 91003, Evry cedex, France

**Abstract**

A compositionally-graded steel composed of martensite with 0.4%C on the centre and bainite with 0.1%C on the surface was manufactured by partial decarburization. It is reported that the as quenched material can be cold rolled up to an equivalent strain of 1.5 without cracks. The mechanical properties of the cold-rolled material exhibits up to 2.4 GPa strength and ductility. A simple mechanical model is developed to predict the stress state after rolling of the graded structure explaining the good ductility of the present high strength materials.

# 1. Introduction

The challenge of developing high strength steel with reasonable ductility is the subject of numerous studies in the metallurgical field. The most effective way of producing high strength steel is to increase the carbon content. This has the obvious drawback of greatly reducing the ductility of the steel. An alternative approach is severe plastic deformation (SPD) to obtain ultra fine grains (UFG). The level of strength of this material is high (up to 2.5 GPa) but the ductility in tension and the work hardening are limited [1]. Recently different studies [2-7] showed the possibility to obtain UFG by cold rolling martensite and/or annealing at warm temperature. For this material it is possible to obtain high ductility (13.2% elongation) with relative high strength (1300 MPa) but the work hardening is very low. In addition to producing UFG microstructures, the above studies provide a valuable insight into the mechanical behaviour of martensite and cold-rolled martensite in terms of strength, uniform elongation and work-hardening. Unfortunately this technique is limited by the ability to cold roll martensite to high levels of deformation. In previous studies [6-7] the carbon content of the martensite is not higher than 0.25%C and the maximum equivalent strain is not more than 1. This is once again due to the fact that for higher carbon content martensite is brittle. In this study we have chosen to work with compositionally graded steel [8] which consisted of martensite with 0.4%C in the centre and bainite with 0.1%C on the surface. The aim is to show that it is possible to keep the high strength of martensite while increasing the ductility by introducing a soft phase at the surface of the material. The improved properties of graded martensite will be illustrated using the examples of as-quenched decarburized martensite and cold-rolled decarburized martensite.

# 2. Experimental procedure

The material used in this study is a commercial 4340M steel whose chemical composition (in wt%) is Fe - 0.39%C - 0.66%Mn - 0.34%Mo - 0.72%Cr - 1.75%Ni - 1.66%Si - 0.1%Al - 0.06%V - 0.003%P. The initial sample thickness is 3 mm. The compositionally-graded steel was obtained using a partial decarburization heat-treatment under a $CO/CO_2$ gas mixture at a 1075°C for a duration of 90 min, followed by oil quenching. The $CO/CO_2$ gas ratio was set to 28.7 in order to obtain a surface concentration of 0.1 wt% C as calculated from the TCFE 2 database of Thermocalc [9]. The duration of the decarburization treatment was chosen to leave a region of approximately 1 mm at the centre of the specimen where the concentration of C is still 0.4%C. A conventional laboratory-scale cold-rolling mill was used to cold roll the as-quenched decarburized steel. At each step the thickness was reduced by 0.04 mm. It was possible to deform the material up to an equivalent strain of 1.5 without any cracking. In what follows, a final thickness of 1.15 mm corresponding to an equivalent strain of 1 was used obtained after 46 rolling passes. The final width of the sample was 4.15 mm whereas 2.8 mm.

The hardness was measured across the thickness of each material by Vickers micro hardness with a load of 0.5 kg applied for 10s. Tensile testing was performed on flat specimens with a gauge length of 15 mm. An Instron 4320 frame with a 100 kN load cell was used. The specimens were pulled to fracture at a constant cross-head speed of 1 mm/minute.

**3. Results and discussion**

**3.1 Hardness profile**

The hardness profiles are presented in the Fig. 1. For comparison, the hardness levels that would be obtained using (homogenous) as-quenched martensite of various carbon contents are included in the Fig. 1. The parabolic hardness profile is observed due to the carbon concentration gradient obtained by decarburization. In the case of the as-quenched sample the difference in hardness between the surface and the core is about 280 HV. The microstructure gradually changed from the surface to the centre with no clear interface between the two.

The parabolic hardness profile is maintained after cold rolling. The contrast between centre and surface is close to 350 HV in the specimen rolled to an equivalent strain of 1. The hardness at the centre is about 900 HV which corresponds to the hardness of homogenous martensite with 1.0 wt%C [10].

The large deformation of the graded material implies that even the martensite at the centre is fully plastic. Examination of the microstructure of the martensite at the centre of the specimen in both the as-quenched and the cold-rolled condition by scanning electron microscopy showed no cracks or voids. This confirms that it is possible to cold-roll the 0.4%C martensitic core up to an equivalent strain of 1 without creating any damage.

A first application of the cold rolling of the graded martensitic steel is the evaluation of the work hardening behaviour of high carbon martensite at high strains. This information is lacking in the literature [2-6]. In the following analysis, the graded material is divided into layers and iso-strain conditions are assumed. Thus, the strain in each layer can be calculated from the total thickness of cold rolled graded steel. The hardness of each layer is then measured as a function of the rolling strain. The initial carbon concentration profile of the graded steel was estimated from the following equation which relates the carbon content $C_m$ to the hardness of the as-quenched martensite [10]:

$$Hvm_{AQ} = 150 + (940 - 150)(1 - \exp(-2{,}7C_m))  \quad\quad\quad \text{Eq. 1}$$

Fig. 2a shows the calculated carbon concentration profile along the thickness direction for the decarburised as well as the decarburized and cold-rolled (strain=1) 4340M. The carbon content at the surface determined using Eq. 1 is more than 0.1%C. This is due to the fact that Eq. 1 was defined for martensite only whereas bainite is observed near the surface. For this reason it was decided to focus further analysis on carbon contents of at least 0.3%C. By varying the equivalent strain obtained by cold rolling it is possible to show the evolution of the hardness of the martensite for carbon contents ranging from 0.3 to 0.4%C (Fig. 2b). For strain more than 0.5 the work hardening rate becomes approximately linear for all the carbon content observed.

### 3.2 Tensile properties

The mechanical properties of the both materials were assessed by tensile tests performed at a strain rate of 0.005 sec$^{-1}$ and the data plotted in the form of engineering stress–strain curves. Curves for monolithic martensite with different carbon contents [10] are included, for comparison, in Fig. 3.

The characteristic values of these curves are reported in the Table 1, together with the uniform elongation measurements. It is interesting to point out that for approximately the same uniform elongation; the UTS of the as-quenched decarburized sample are 13% greater than that of the homogenous martensite with 0.3%C. The comparison with the 0.4%C homogenous martensite shows that for a decrease of 5% in UTS the graded materials achieved at 39% increase in uniform elongation. Because of the very high strength of the cold-rolled decarburized steel it is more appropriate to compare it to a higher carbon martensite with a similar strength: the cold-rolled graded material and the 0.5%C martensite have a similar strength (in term of hardness). Compared to the 0.5%C martensite exhibits an extended elasto-plastic transition and a higher level of ductility. For both the graded and cold-rolled graded material, the work-hardening rate is sustained at a high level characteristic of high carbon martensite while achieving improved ductility compared to other microstructures with similar strength.

In order to compare the present results with the previous works [6, 7] on cold-rolled martensite, the maximum strength versus equivalent Von Mises strain produced by cold-rolling are reported in Fig. 4. The superior combination of strength and ductility in the graded material is clearly obvious.

From these tensile specimens the ductility at fracture was assessed from measurements of the final reduction in area (RA %). These values as well as the ones known for martensite with different carbon contents are given in Table 1. Comparison of as-quenched decarburized steel with monolithic 0.4%C martensite shows that the effect of the gradient is dramatic; in fact the RA is increased by 400% whereas the UTS decreased only by 5%. The RA% of the cold-rolled decarburized material is higher than that of the homogenous 0.4%C.

The fracture surfaces were observed by SEM (Fig. 5). Two distinct zones are visible for both graded materials; ductile fracture on the outer surface and brittle fracture at the centre. The fracture at the centre is an intergranular fracture corresponding to the prior austenitic grain boundaries. This was checked by comparing the scale of the primary austenite grain with the characteristic scale of fracture identified on the SEM observations. It is interesting to observe that after cold rolling the characteristic scale of the fracture decreased in a way that approximately scales with the cold-rolling reduction. A second, smaller, characteristic scale is observed on the fracture surface as shown in Fig. 5, but its significance is not well understood for the moment.

The limit of the ductile outer zone appears to correspond to a carbon content of about 0.2%C. The ductile outer material is bainite or autotempered martensite which explains the apparition of ductile fracture. The dimple/void size in the ductile outer layer decrease from about 5µm for the as-quenched material to less than 2µm for the cold rolled material. This can be explained by the decrease of the toughness of the material after cold rolling.

**3.3 Modelling of the stress state during cold rolling of the graded steel**

One of the attractive properties of the graded martensitic steel showed in this study is that it is possible to cold roll martensite with 0.4%C to equivalent strains of more than 1 without cracking. This phenomenon could simply be explained in terms of the outer soft layer protecting the core by preventing the propagation of surface cracks. Fig. 5 clearly shows the ductile nature of the fracture in the outer surface layer. Another interesting property of the cold-rolled graded material is its reasonable ductility. It is therefore of interest to make the analysis of the stress distribution during the cold rolling as this would provide further insight into these two phenomena. Analytical analysis of the mechanical behaviour of the cold rolling of a three layer materials composed of an internal hard layer and two soft materials has already been presented in the literature [11]. The aim was to show that sandwiching the hard core between two softer layers (e.g. by cladding) can reduce the pressure applied during cold rolling in a rigid plastic material under plane strain conditions.

In the following analysis the distribution of stress in graded materials along the thickness direction is examined. The x-axis and y-axes are defined along the rolling and thickness directions, respectively, with the origin being at the centre of the material. The symbols, h, p and $\sigma_x$ represent, respectively, the thickness of the graded material, the pressure applied along y-axis to the graded material during cold rolling (positive) and the stress applied on the material along the x-direction (positive for compression). The various layers of the material are considered perfectly plastic. The concentration gradient in the y-direction leads to a flow stress gradient. As a very good approximation, the flow stress variation in the y-direction, $\sigma_f(y)$, is assumed to be expressed as:

$$\sigma_f(y) = a - b \cdot y^q \qquad \text{Eq. 2}$$

where a is the flow stress at the centre, b and q are constants which could be determined from Fig. 2. The cold rolling can be considered as plain-strain compression. In these conditions the Tresca yield criterion is written as:

$$p - \sigma_x = 2.\sigma_f(y) \qquad \text{Eq. 3}$$

The zero force boundary condition applied along x axis gives:

$$\int_o^{h/2} \sigma_x .dy = 0 \qquad \text{Eq. 4}$$

By substituting Eqs. 2 and 3 into Eq. 4 and integrating one obtains:

$$p = 2\left(a - 2b.\frac{h^q}{(q+1).2^{(q+1)}}\right) \qquad \text{Eq. 5}$$

and finally by substituting p and $\sigma_f(y)$ into Eq. 3 the expression of $\sigma_x$ is expressed as:

$$\sigma_x = 2b\left(y^q - 2\frac{h^q}{(q+1).2^{(q+1)}}\right) \qquad \text{Eq. 6}$$

Considering b>0 allowed to show that the stress at the centre is negative which according to our sign convention means that the centre is in tension whereas the surface is in compression. The transition is defined as the point for which $\sigma_x$ is equal to 0. This occurs at:

$$y = \left( \frac{2}{(q+1) \cdot 2^{(q+1)}} \right)^{1/q} h \qquad \text{Eq. 7}$$

Considering the hardness profile of the cold rolled graded steel 4340M at an equivalent strain of 1 the distribution of flow stress along the thickness can be described. Using the above analysis it is possible to evaluate the distribution of $\sigma_x$ along the thickness of the graded material as shown in Fig. 6. For this application a=2700 MPa, b=1050 N and q=2. These parameters were chosen to fit approximately with the hardness measurements presented in Fig. 2.

The results show that the material at the centre is under tension ($\sigma_x < 0$); this region represents approximately 30% of the total thickness of the material. The remaining material is under compression. In fact, the compression stress at the surface is very high.

Based on the above analysis two interesting points have to be discussed. Firstly it is clearly shown that the hardest material at the core is under tension. This can not explain the fact that the technique of cold rolling the graded steel allows to cold rolled 0.4%C martensite. It therefore seems that the soft layer in compression introduced on the surface is a more plausible explanation for the improved cold-workability of the graded material. Secondly this analysis helps to understand the behaviour of the cold rolled graded steels and understand that the level of ductility for such strength material. Since the bigger part of the material after cold rolling is under compression, the effective effect of the soft phase at the surface is improved.

## 4. Conclusions

In this study compositionally-graded steel was manufactured by partial decarburization. The microstructure is composed of martensite with 0.4%C on the centre and bainite with 0.1%C on the surface. From this we showed some of the interests of the graded steel.

- The as-quenched material can be cold rolled up to an equivalent strain of 1.5 without cracks and then with this kind of material it is possible to cold rolled a martensite with 0.4%C up to an equivalent strain of 1.5

- Thanks to the possibility to cold rolled the graded structures at high equivalent strain the work hardening behaviour of martensite with 0.3, 0.35, 0.4 %C for equivalent strains up to 1.3 was determined. It appears that the work hardening become approximately linear for deformation high than 0.5. This means that we do not observe saturation in work hardening of the martensite for high deformation.

- The mechanical properties of the cold-rolled material at an equivalent strain of 1 were significantly better in comparison with a 0,4% C homogeneous martensitic steel. The graded steel exhibits and extended elasto-plastic behavior and a tensile strength of 2.5 GPa. The fracture surface exhibits a ductile zone near the surface and a brittle zone near the centre.

- A simple model to evaluate the stress distribution after cold rolling of the graded steel along the thickness was developed. It shows clearly that the soft outer zone is in compression which can prevent the propagation of surface cracks into the hard core.

**Acknowledgement:**

Authors gratefully acknowledge the financial support of the Natural Science and Engineering Research Council (NSERC) of Canada.

[1] O. Bouaziz, *Mater. Sci. Forum*, **2009**, 633-634, 205

[2] N. Tsuji, R. Ueji, Y. Minamino, Y. Saito, *Scripta Mater.* **2002,** 46, 305

[3] R. Ueji, N. Tsuji, Y. Minamino, Y. Koizumi, *Acta Mater.* **2002**, 50, 4177

[4] X. Zhao, T.F. Jing, Y.W. Gao, G.Y. Qiao, J.F. Zhou, W. Wang *Mater. Sci. Eng. A* **2005**, 397, 117

[5] J. Tianfu, G. Yuwei, Q. Guiying, L. Qun, W. Tiansheng, W. Wei, X. Furen, C. Dayong, S. Xinyu, Z. Xin *Mater. Sci. Eng. A* **2006**, 432, 216

[6] R. Ueji, N. Tsuji, Y. Minamino, Y. Koizumi, *Sci. and Tech. of Adv. Mater.* **2004**, 5, 153

[7] J.C. Lee, U.G. Kang, C. S. Oh, S.J. Kim, W. J. Nam, *Mater. Sci. Forum*, **2009**, 654-656, 218

[8] B. Chéhab, H. Zurob, D. Embury, O. Bouaziz, Y. Bréchet, *Adv. Eng. Mater.* **2009**,12, 992

[9] B. Sundman, B. Jansson, JO. Andersson, *CALPHAD 9* **1985**, 153

[10] G. Krauss, *Mater. Sci. and Eng. A* **1999**, 273-275, 40

[11] A. A. Afonja, D. H. Sansome, *Int. J. mech. Sci*. **1973**, 15, 1-14

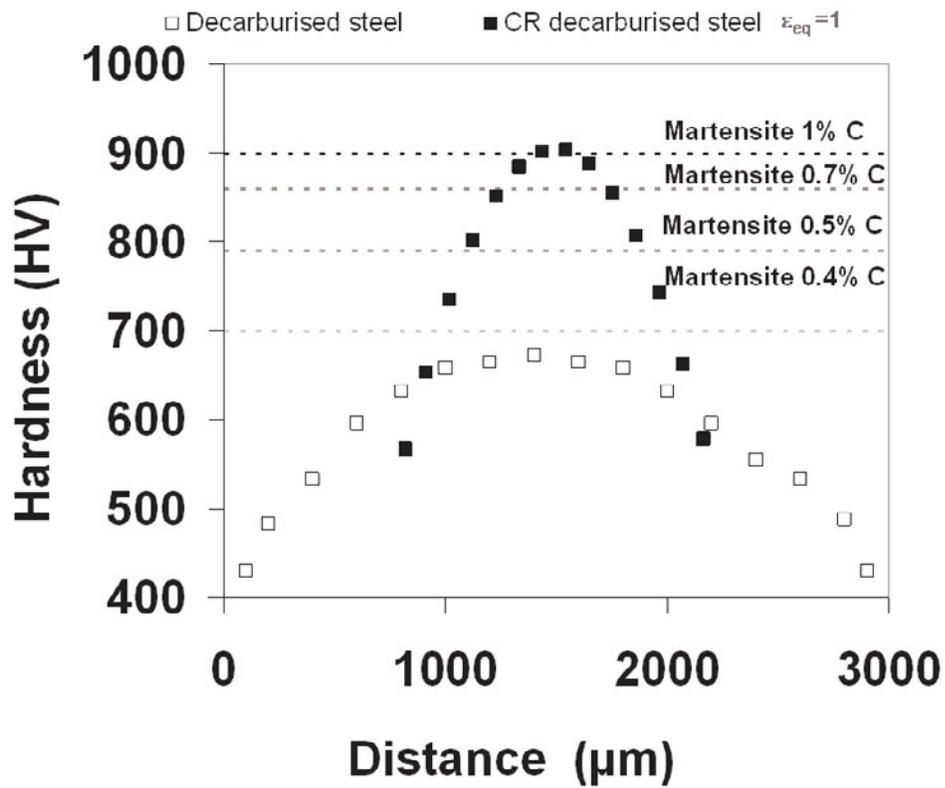

**Fig. 1 :** Hardness profile of partially decarburised steel (4340M) and cold rolled decarburised steel (4340M) at an equivalent strain of 1. Comparison is done with hardness of homogeneous martensite with different carbon content (data from [10])

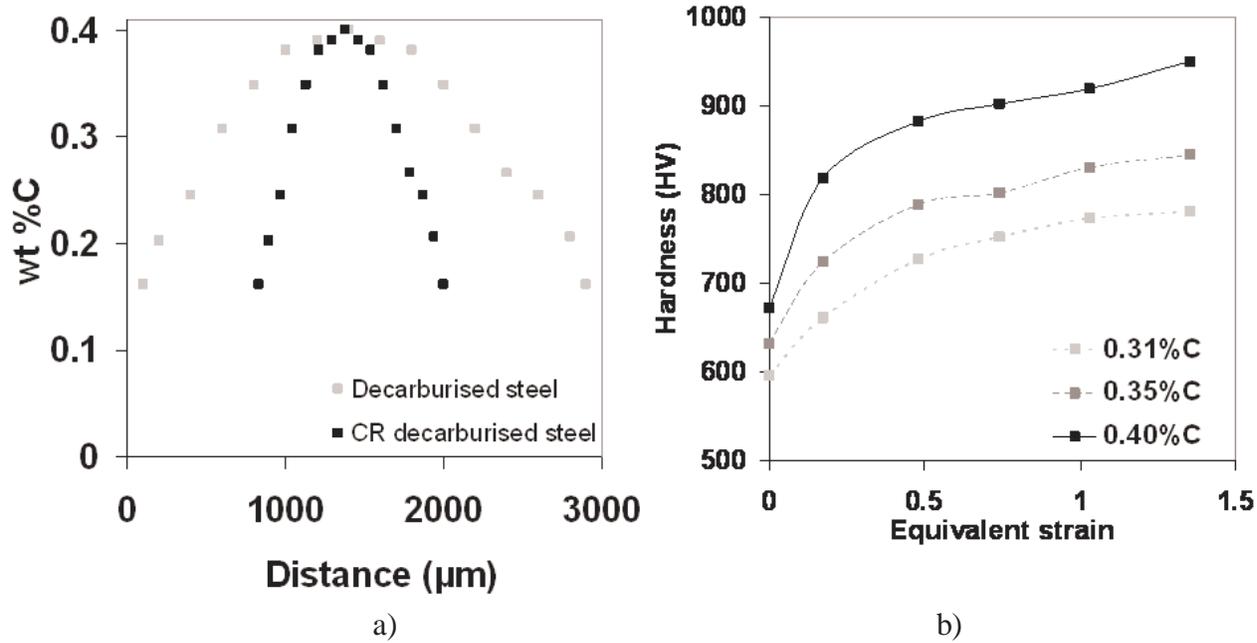

a)             b)

**Fig. 2:** a) Carbon content profile along the thickness deduced from hardness measurements for partially decarburised steel (4340M) and cold rolled decarburised steel (4340M) at an equivalent strain of 1. b) Hardness evolution with equivalent prestrain obtained by cold rolling for martensite with 0.31, 0.35 and 0.4 %C.

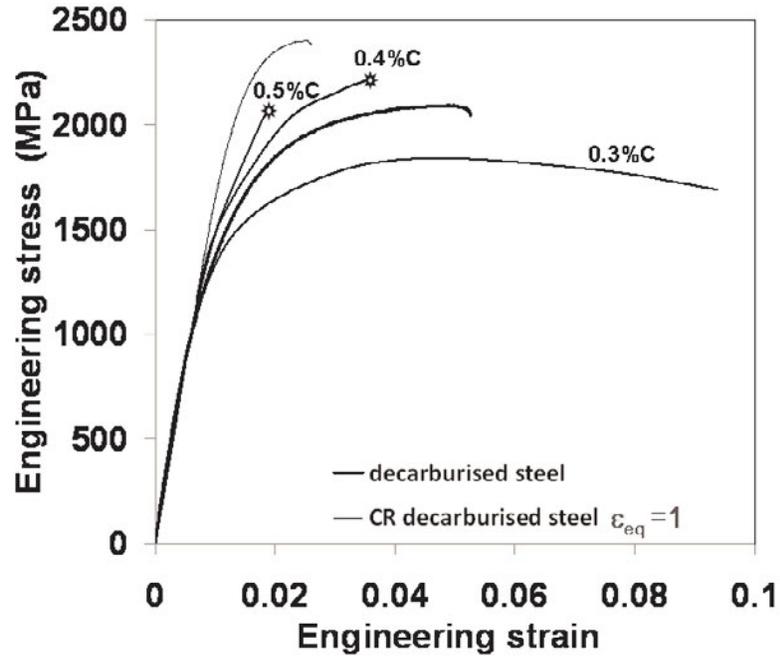

**Fig. 3:** Engineering stress-strain curves of partially decarburised steel (4340M) and cold rolled decarburised steel (4340M) at an equivalent strain of 1. Comparison is done with curves of homogeneous martensite with different carbon content (data from [10]). * means that breaking occurs without necking.

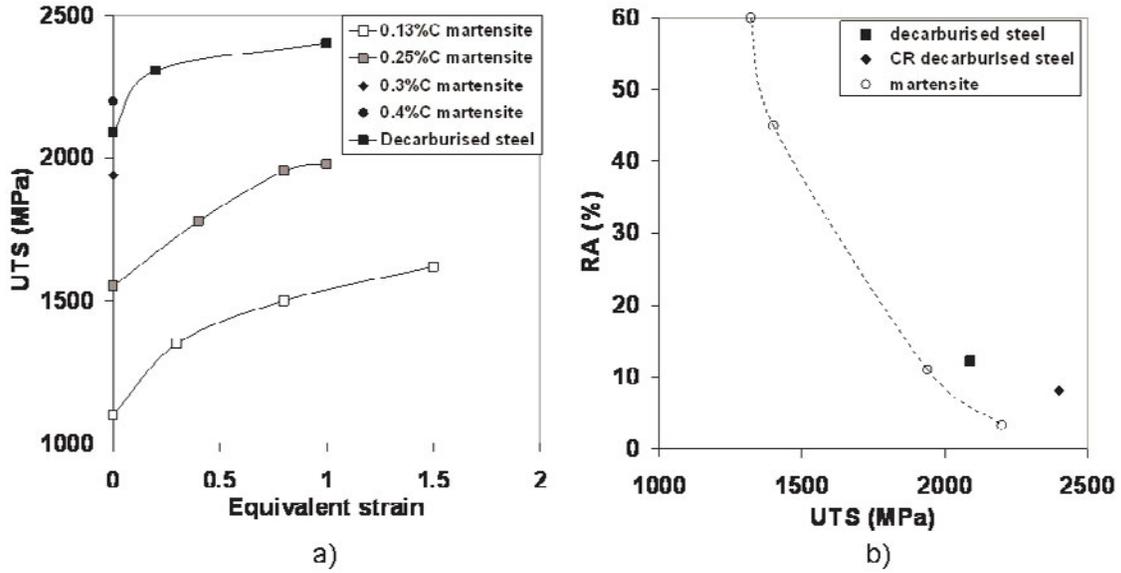

**Fig. 4 :** a) Ultimate Tensile Strength versus equivalent strain obtained by cold rolling for partially decarburised steel (4340M) compared with 0.13%C martensite [6] and 0.25%C martensite [7]. The UTS of non deformed martensite with different carbon content is added. b) Reduction Area (RA%) versus UTS for partially decarburised steel (4340M) and cold rolled decarburised steel (4340M) at an equivalent strain of 1 compared with values for different homogeneous martensite with different carbon content (data from [10])

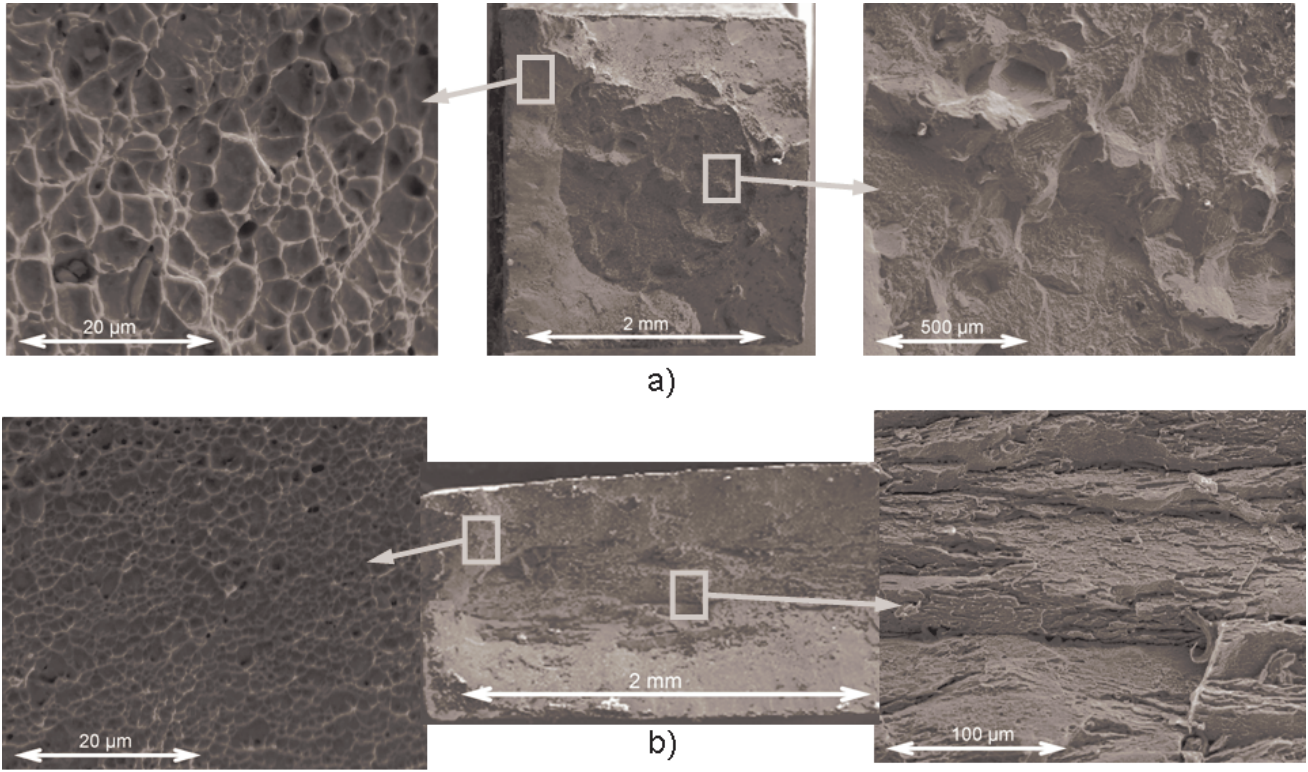

**Fig. 5 :** SEM observation of fracture surface of a) decarburized steel (4340M) and b) cold rolled decarburized steel (4340M) at an equivalent strain of 1.

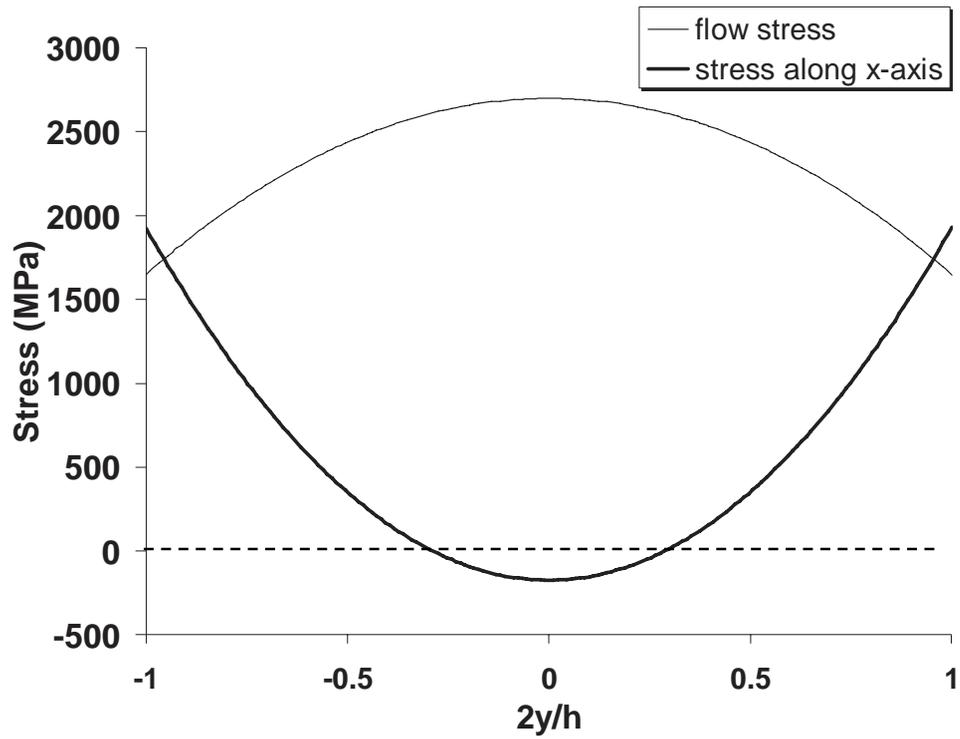

**Fig. 6 : Prediction of the stress state along the rolling direction considering the gradient of property observed on the cold rolled decarburized steel (4340M) at an equivalent strain of 1 to know the flow stress. In this Figure we adopted the sign convention in which tensile stress in the x-direction is negative.**

|  | 0.3%C Martensite | 0.4%C Martensite | 0.5%C Martensite | As quenched decarburized | CR decarburized |
|---|---|---|---|---|---|
| $\varepsilon_u$ (%) | 4.8 | 3.5* | 1.9* | 4.9 | 2.5 |
| UTS | 1840 | 2210 | 2060 | 2090 | 2400 |
| RA% | 11,2 | 3.3 |  | 12.2 | 8.1 |

**Tab. 1 : Mechanical characteristics values ($\varepsilon_u$: uniform elongation, UTS: ultimate tensile strength, RA%: reduction area) from tensile test for decarburized steel (4340M) and cold rolled decarburized steel (4340M) at an equivalent of 1 compared with value for homogeneous martensite with different carbon content (data from [10]). * means fracture without necking.**